\documentclass[a4paper,12pt]{article}
\usepackage{amsfonts}
\def\eqsecnum{%
        \@addtoreset{equation}{section} %
        \def\theequation{\arabic{section}.\arabic{equation}}%
}

\newcommand{\fr}[2]{\frac{#1}{#2}}
\newcommand{\frp}[2]{\frac{\partial {#1}}{\partial {#2}}}

\newcommand{\re}[1]{(\ref{#1})}
\newcommand{\beq}{\begin{equation}}
\newcommand{\beql}[1]{\begin{equation}\label{#1}}
\newcommand{\eeq}{\end{equation}}
\newcommand{\pl}{\partial}
\newcommand{\nn}{\nonumber}
\newcommand{\q}{\quad}
\newcommand{\qq}{\qquad}
\newcommand{\bb}[1]{{\mathbb {#1}}}
\newcommand{\bm}[1]{{\mathbf {#1}}}
\newcommand{\bg}[1]{\mbox{\boldmath${#1}$}}
\newcommand{\ca}[1]{{\cal {#1}}}
\newcommand{\go}[1]{{\mathfrak {#1}}}
\newcommand{\ve}{\varepsilon}

\title{Classical relativistic systems of charged particles\\
in the front form of dynamics\\
and the Liouville equation}
\author{A.~Nazarenko, V.~Tretyak\\[5pt]
\small Institute for Condensed Matter Physics\\
\small of the National Academy of Science of Ukraine\\
\small1 Svientsitskyi Str., Lviv, 79011, Ukraine,\\ 
\small e-mail: tretyak@icmp.lviv.ua}
\date{}
\begin{document}
\maketitle
\renewcommand{\abstractname}{\ }
\begin{abstract}
Classical relativistic system of point particles coupled with an
electromagnetic field is considered in the three-dimensional
representation. The gauge freedom connected with the chronometrical
invariance of the four-dimensional description is reduced by use of the
geometrical concept of the forms of relativistic dynamics. The remainder
gauge degrees of freedom of the electromagnetic potential are analysed
within the framework of Dirac's constrained Hamiltonian mechanics in the
front form of dynamics. The results are implemented to the problems of
relativistic statistical mechanics. Based on the corresponding
Liouville equation the classical partition function of the system is
written down in a gauge-invariant manner and an integration over field
variables is performed.


\end{abstract}

\newpage
\section{Introduction}

All fundamental interactions in physics have got gauge nature and demand the
use of singular Lagrangians. An adequate method for dealing with such
systems was developed by Dirac (see \cite{dirac,SudM,Sund}). His
approach, i.e., constrained Hamiltonian mechanics, has been elaborated in many
directions \cite{mmt,got} and usefully applied to the problems of
quantum field theory (see, e.g., \cite {fad,slav}).
But an application of this technique to the relativistic statistical
mechanics \cite{hor,drv} seems rather obscure, although, for example,
the thermo field formulation of the condensed matter theory \cite{tfd}
admits a variational approach with the same gauge structure as in quantum
electrodynamics and in related theories. We only note the paper \cite{mil},
where the use of Faddeev' measure \cite{fad} was suggested for the
construction of the relativistic partition function of directly
interacting particle system described by means of the constrained
Hamiltonian formalism.

The present paper is concerned with the classical relativistic system of
point charges coupled with an electromagnetic field. The first attempt
to analyse the constraint contents of such a system was made by Dirac
\cite{dir}. The corresponding action functional in the four-dimensional
representation has got two kinds of the gauge freedom. The first is
connected with arbitrariness in the parametrisation of particle world
lines (chronometrical invariance); the second is generated by the proper
gauge transformations of electromagnetic potentials. We reduce the gauge
freedom of the first kind by means of the farther great Dirac's
invention, namely, the concept of the forms of relativistic dynamics
\cite{dir49}. Here we use the notion of the form of dynamics as denoting
the description of relativistic system, which corresponds to a given
global simultaneity relation defined by means of some foliation of the
Minkowski space by space-time or isotropic hypersurfaces. Moreover, we
consider only the case when the corresponding simultaneity relation is 
independent of particle or field configurations (cf.\ \cite{lus}). Using 
Dirac's constraint formalism, we try to isolate the gauge degrees of 
freedom and to formulate the statistical description of the system in a 
gauge-invariant manner.

The general features of the formalism were presented in \cite{dn} within
the instant form of dynamics. It will be our purpose in the present
paper to explore the further possibilities connected with other forms of
relativistic dynamics; in particular, we shall consider the front form
description of the charged particle system.

The paper is organised as follows. In Sec.~2 we will first establish the
structure of Hamiltonian description of the system of particles plus field
in an arbitrary form of relativistic dynamics. Our results are then
applied to the front form of dynamics. In Sec.~3 we perform the analysis
of the corresponding constraints. Section 4 contains the elimination of the
gauge degrees of freedom by suitable canonical transformation. Sections
5 and 6 are devoted to the application in relativistic statistical
mechanics. We formulate a Liouville equation for distribution function
in the front form of dynamics and write down its equilibrium solution
corresponding to classical Gibbs ensemble accounting particle and field
degrees of freedom. Peculiarity of the front form of dynamics allow us
to perform an integration over field variables in a relativistic
partition function. Some calculations of a purely technical nature are
collected in appendices.

\section{Charged particle system in an arbitrary form of dynamics}
\setcounter{equation}{0}

We shall consider a system of $N$ charged particles, which is
described by their (time-like) world lines in the Minkowski space-time%
\footnote{The Minkowski space-time $\bb M_4$ is endowed with a metric
$\| \eta_{\mu \nu } \| = \mathrm {diag} (-1,1,1,1)$. The Greek indices
$\mu,\nu,\ldots$ run from 0 to 3; the Latin indices from the middle of
alphabet, $i,j,k,\ldots$, run from 1 to 3 and both types of indices
are subject of the summation convention. The Latin indices from the
beginning of alphabet, $a,b$, label the particles and run from 1 to $N$.
The sum over such indices is pointed explicitly. The velocity of light
and the Planck constant $\hbar$ are equal to unity.}
\eqsecnum
\beql{2.1}
\gamma_a: \bb R \mapsto \bb M_4,\quad \tau \mapsto x^{\mu}_a(\tau).
\eeq
An interaction between charges is assumed to be mediated by an
electromagnetic field $F=dA$ with the electromagnetic potential 1-form
\beql{2.2}
A=\tilde A_{\mu}(x)dx^{\mu};
\eeq
\beql{2.3}
F=\fr12\tilde F_{\mu\nu}(x)dx^{\mu}\wedge dx^{\nu}, \quad
\tilde F_{\mu\nu}(x)=\pl_{\mu}\tilde A_{\nu}-\pl_{\nu}\tilde A_{\mu},
\eeq
$\pl_{\nu}\equiv\pl /\pl{x^{\nu}}$. The dynamical properties of such
a system are completely determined by the following action functional
\cite{ll,ror,parrot}
\beql{2.4}
S = \sum_{a=1}^{N}\int d\tau_a\left \{-m_a\sqrt{-u^2_a(\tau_a)}+
e_au^{\nu}_a(\tau_a)\tilde A_{\nu}[x_a(\tau_a)]\right \}
-\frac{1}{16\pi}\int d^4x\tilde F_{\mu \nu }(x)\tilde F^{\mu \nu }(x),
\eeq
where $m_a$ and $e_a$ denote the mass and the charge of particle $a$,
respectively, $u^{\mu}_a(\tau_a)=dx^{\mu}_a(\tau_a)/d\tau_a$.
We are interested in constructing the Hamiltonian description of the
system in a given form of relativistic dynamics. The particle and field
degrees of freedom will be treated on equal level.

The form of relativistic dynamics in its geometrical definition is
specified by the foliation $\Sigma=\{\Sigma_t~|~t\in\bb R\}$ of the
Minkowski space-time with space-like or isotropic hypersurfaces
\beql{2.5}
\Sigma _t=\{x\in\bb M_4~|~\sigma (x)=t\};
\eeq
\beql{2.6}
\eta^{\mu\nu}(\pl_{\mu}\sigma)(\pl_{\nu}\sigma)\leq 0.
\eeq
(see \cite{gkt83,gkt89,dst}).
As it follows from condition \re{2.6}, $\pl_0\sigma\not=0$ and therefore
the hypersurface equation \re{2.5} can be solved with respect to $x^0$
in the form:
\beql{2.7}
x^0=\varphi (t,\bm x), \q \bm x=(x^1,x^2,x^3).
\eeq
For definiteness we put
\beql{2.8}
\pl_0\sigma>0 \q{\rm and}\q
\frp{\varphi (t,\bm x)}t\equiv\varphi_t(t,\bm x)>0.
\eeq
Making use of the identities
\beql{2.7a}
\left. \pl_0\sigma\right |_{\Sigma_t}=\varphi_t^{-1}, \qq
\left. \pl_i\sigma\right |_{\Sigma_t}=-\varphi_i\varphi_t^{-1}, \qq
\varphi_i\equiv\pl\varphi/\pl x^i,
\eeq
it is easy to see that condition \re{2.6} implies the following
inequality for the function $\varphi$:
\beql{2.7b}
\bg {\varphi}^2\leq 1, \qq
\bg {\varphi}=\left(\frp{\varphi}{x^1}, \frp{\varphi}{x^2},
\frp{\varphi}{x^3}\right).
\eeq

Then one can consider the particle world lines $\gamma_a$ as being
determined by a set of points $x_a(t)=\gamma _a\bigcap\Sigma _t$
of intersections with the elements of the foliation $\Sigma$. The
parametrical equation of the world line in a given form of dynamics
is 
\beql{2.9} 
x^0=x_a^0(t)=\varphi (t,\bm x_a(t))\equiv\varphi_a, \qq x^i=x_a^i(t).
\eeq
The variable $t$ serves as a common evolution parameter of the particle
system. Therefore, the choice of the form of dynamics is equivalent to
the fixing of the parameters $\tau_a=t$ of the particle world lines in
the reparametrization-invariant action \re{2.4}.

On the other hand, the foliation $\Sigma$ specifies a certain $3+1$
splitting of the Minkowski space-time
$\bb M_4\cong\bb R\times\Sigma_0$, because by the definition of a
foliation all hypersurfaces $\Sigma_t$ for different $t$ are
diffeomorphic to $\Sigma_0$. That splitting,
$f: \bb M_4\to\bb R\times\Sigma_0$, is really determined by \re{2.7}:
\beql{2.10}
f: (x^0, \bm x)\mapsto (\varphi(t,\bm x),\bm x).
\eeq

Accounting that the determinant of the transformation \re{2.10} is
\beq
\fr{\pl(x^0,\bm x)}{\pl(t,\bm x)}=\varphi_t(t,\bm x),
\eeq
one can immediately rewrite the action functional \re{2.4} into the form
\beql{2.12}
S=\int dtL
\eeq
with the Lagrangian function
\beql{2.13}
L=\sum_{a=1}^{N}\left\{-m_a\sqrt{(D\varphi_a)^2-\bm v_a^2}+
e_a\left[A_0(t,\bm x_a)D\varphi_a+v_a^iA_i(t,\bm x_a)\right]\right\}+
\int d^3x\ca L;
\eeq
\beql{2.14}
\ca L=-\fr1{16\pi}\varphi_t(t,\bm x)F_{\mu\nu}F^{\mu\nu}.
\eeq
Here $\bm v_a=d\bm x_a/dt$, $D=d/dt=\pl/\pl t+\sum_av_a^i\pl/\pl x_a^i$,
and $A_{\mu}=\tilde A_{\mu}\circ f^{-1}$,
$F_{\mu\nu}=\tilde F_{\mu\nu}\circ f^{-1}$.
The dynamical variables of our variational problem will be the
functions $\bm x_a(t)$, $A^{\mu}(t,\bm x)$ and their first order
derivatives with respect to the evolution parameter, $\bm v_a(t)$ and
$A^{\mu}{}_{,t}(t,\bm x)$. Using the obvious relations
\beql{2.15}
\pl_0\tilde A_{\mu}=A_{\mu,t}\varphi_t^{-1}, \qq
\pl_i\tilde A_{\mu}=A_{\mu,i}-A_{\mu,t}\varphi_i\varphi_t^{-1},
\eeq
we find the following expression for the field Lagrangian density:
\beql{2.16}
\ca L= \fr{\varphi_t}{8\pi}\left(e_ie^i-h_ih^i\right),
\eeq
where
\beql{2.17}
e_i=-A_{0,i}+\left(A_{i,t}+A_{0,t}\varphi_i\right)\varphi_t^{-1}, \q
h_i=\ve_{ijk}\left(A^{k,j}-A^k{}_t\varphi^j\varphi_t^{-1}\right).
\eeq

The canonical momenta of our problem are given by
\beql{2.18}
p_{ai}=\frp L{v^i_a}=
\fr{m_a(v_{ai}-\varphi_{ai}D\varphi_a)}{\sqrt{(D\varphi_a)^2-\bm v_a^2}}+
e_a\left[A_i(t,\bm x_a)+A_0(t,\bm x_a)\varphi_{ai}\right],
\eeq
\beql{2.19}
{\mit \Pi}_i(t,\bm x)=\frp{\ca L}{A^i{}_{,t}}=
\fr{1}{4\pi}\left(e_i-\ve_{ijk}\varphi^jh^k\right),
\eeq
\beql{2.20}
{\mit \Pi}(t,\bm x)=-\frp{\ca L}{A_{0,t}}=
-\fr{1}{4\pi}e_i\varphi^i.
\eeq
The basic Poisson brackets are
\begin{eqnarray}
&&
\{x^i_a, p_{bj}\}=\delta_{ab}\delta^i_j, \nn \\
&&
\{A_i(t,\bm x),{\mit\Pi}_j(t,\bm y)\} =
\delta_{ij}\delta^3(\bm x-\bm y), \label{2.21} \\
&&
\{A_0(t,\bm x),{\mit\Pi}(t,\bm y)\}=-\delta^3(\bm x-\bm y);\nn
\end{eqnarray}
all other brackets vanish. Equations \re{2.19} and \re{2.20} imply the
primary constraint
\beql{2.22}
\xi\equiv{\mit\Pi}+{\mit\Pi}^i\varphi_i=0.
\eeq

In the isotropic forms of dynamics, however, some additional primary
constraints may occur. To see that, we rewrite equation \re{2.19} into
the form
\beq
{\mit \Pi}_i(t,\bm x)=
\fr{1}{4\pi}\left(\go S_i+\beta_{ij}A^j{}_{,t}\varphi_t^{-1}
+A_{0,t}\varphi_i\varphi_t^{-1}\right),
\eeq
where $\go S_i$ does not contain the derivatives with respect to the
evolution parameter $t$,
\beq
\go S_i=-A_{0,i}-\varphi^jh_{ij}; \qq 
h_{ij}\equiv A_{j,i}-A_{i,j},
\eeq
and
\beq
\beta_{ij}=(1-\bg {\varphi}^2)\delta_{ij}+\varphi_i\varphi_j.
\eeq
Upon using the identity $\beta_{ij}\varphi^j=\varphi_i$, one obtains 
equation
\beql{2.25}
\beta_{ij}\left(A^j{}_{,t}+A_{0,t}\varphi^j\right)
=\varphi_t\left(4\pi{\mit\Pi}_i-\go S_i\right).
\eeq
The matrix $\beta=\|\beta_{ij}\|$ has determinant
$\det\beta=(1-\bg {\varphi}^2)^2$.

In the case of space-like forms of dynamics ($\bg {\varphi}^2<1$) the
matrix is nonsingular and possesses an inverse
\beql{2.26}
\gamma_{ij}=\fr{\delta_{ij}-\varphi_i\varphi_j}{1-\bg {\varphi}^2}, \qq
\beta^{ij}\gamma_{jk}=\delta^i_k,
\eeq
so that equation \re{2.25} can be solved as
\beql{2.27}
A_{i,t}+A_{0,t}\varphi_i=
\gamma_{ij}\varphi_t\left(4\pi{\mit\Pi}^j-\go S^j\right).
\eeq
For isotropic forms of dynamics we have $\beta_{ij}=\varphi_i\varphi_j$
and \re{2.25} implies constraints
\beql{2.28}
\xi_i\equiv\ve_{ijk}\varphi^j\left(4\pi{\mit\Pi}^k-\go S^k\right)=0.
\eeq
Because it holds identically $\varphi^i\xi_i=0$, really only two
additional primary constraints occur.

Next, we consider the canonical Hamiltonian of our system, which is
defined as
\beq
H_{{\rm c}}=\sum_{a=1}^Np_{ai}v_a^i+\int d^3x\left({\mit\Pi}^iA_{i,t}
-{\mit\Pi}A_{0,t}\right)-L.
\eeq
The immediate calculations give
\beq
H_{{\rm c}}=\sum_{a=1}^NH_a(t,\bm x_a,\bm p_a)+\int d^3x\ca H_{{\rm c}}
\eeq
with
\beq
H_a=\fr{m_a\varphi_{at}D\varphi_a}{\sqrt{(D\varphi_a)^2-\bm v_a^2}}-
e_a\varphi_{at}A_0(t,\bm x_a)
\eeq
and
\beql{2.32}
\ca H_{{\rm c}}=\fr{1}{4\pi}\left[\fr12\varphi_t\left(e_ie^i+h_ih^i\right)
+e^iA_{0,i}-\ve_{ijk}A^i{}_{,t}\varphi^jh^k\right].
\eeq
The problem of solution of \re{2.18} with respect to the particle  
velocities is analogous to the case of free particles \cite{prt}.
For completeness we collect the corresponding expressions in Appendix A.
Replacing into \re{a9} $p_i$ by
\beq
k_{ai}\equiv p_{ai}-
e_a\left[A_i(t,\bm x_a)+A_0(t,\bm x_a)\varphi_{ai}\right]
\eeq
yields the one-particle Hamiltonian for the space-like forms of dynamics:
\beq
H_a=\fr{\varphi_{at}}{1-{\bg {\varphi}}_a^2}
\left[\sqrt{(1-\bg {\varphi}_a^2)(m_a^2+\bm k_a^2)+
(k_a^i\varphi_{ai})^2}+k_a^i\varphi_{ai}\right]-
e_a\varphi_{at}A_0(t,\bm x_a).
\eeq
Similarly, for the isotropic forms of dynamics we obtain from \re{a11}
\beq
H_a=-\fr12\varphi_{at}\fr{m_a^2+\bm k_a^2}{k_a^i\varphi_{ai}}-
e_a\varphi_{at}A_0(t,\bm x_a).
\eeq

To express the field Hamiltonian \re{2.32} in the terms of canonical
variables we rewrite it into the form
\beql{2.35}
\ca H_{{\rm c}}=\fr{1}{8\pi}\left\{\varphi_t
\left[\fr12 h_{ij}h^{ij}+A_{0,i}A^{0,i}\right]
+\varphi_t^{-1}\beta_{ij}\left(A^i{}_{,t}+A_{0,t}\varphi^i\right)
\left(A^j{}_{,t}+A_{0,t}\varphi^j\right)\right\}.
\eeq
Then for the space-like forms of dynamics, taking into account
\re{2.26} and  \re{2.27}, we obtain
\beq
\ca H_{{\rm c}}=\fr{1}{8\pi}\varphi_t\left\{
\gamma^{ij}\left(4\pi{\mit\Pi}_i-\go S_i\right)
\left(4\pi{\mit\Pi}_j-\go S_j\right)+
\fr12 h_{ij}h^{ij}+A_{0,i}A^{0,i}\right\}.
\eeq
In the case of isotropic forms of dynamics ($\bg {\varphi}^2=1$),
when $\beta_{ij}=\varphi_i\varphi_j$, \re{2.25} gives
\beq
\varphi_j\left(A^j{}_{,t}+A_{0,t}\varphi^j\right)
=\varphi_t\varphi_j\left(4\pi{\mit\Pi}^j-\go S^j\right),
\eeq
so that \re{2.35} takes the form
\beq
\ca H_{{\rm c}}=\fr{1}{8\pi}\varphi_t\left\{\left[
\varphi^i\left(4\pi{\mit\Pi}_i-\go S_i\right)\right]^2+
\fr12 h_{ij}h^{ij}+A_{0,i}A^{0,i}\right\}.
\eeq
Detailed analysis of the Hamiltonian description for the charged
particle system in the instant form of dynamics was carried out in
\cite{dn} with the application to the classical relativistic statistical
mechanics. In the following sections we shall be concerned with the
front form of dynamics, analyzing additional primary constraints and
ensuring the corresponding Liouville equation.

\section{Constraint analysis in the front form of dynamics}
\setcounter{equation}{0}

Let us consider the family of the forms of dynamics, which is 
given by
\beql{3.1}
x^0 = t + {\bf n x},\ \ {\bf n}^2 = 1.
\eeq
According to \re{2.22}, \re{2.28}, the set of primary constraints is
\begin{eqnarray}
&&
\xi\equiv{\mit\Pi}+{\mit\Pi}^i n_i=0  \label{3.2} \\
&&
\xi_i\equiv\ve_{ijk} n^j\left(4\pi{\mit\Pi}^k+A_0{}^{,k}-
n_l h^{lk}\right)=0. \label{3.3}
\end{eqnarray}
It is easy to see that \re{3.3} is equivalent to
\beql{3.4}
(4\pi{\mit\Pi}^k+A_0{}^{,k})_{\perp}-n_jh^{jk}=0,
\eeq
where we define orthogonal and longitudinal projections of an arbitrary 
3-vector $f^k$ with respect to the vector $n^k$ as
\beql{3.5}
f^k = f^k_{\perp}+f^k_{\parallel}, \ \
f^k_{\parallel} = n^k n_l f^l, \ \
f^k_{\perp} = f^k - n^k n_l f^l.
\eeq

The canonical Hamiltonian of the system is determined by
\begin{eqnarray}
&&
H_{\rm c} = - \sum\limits_{a=1}^{N}\left[
\fr{m_a^2+\bm k_a^2}{2\bm k_a \bm n}+e_aA_0(t,\bm x_a)\right] \nn \\
&&
+\fr{1}{8\pi}\int\left\{\left[
n^i\left(4\pi{\mit\Pi}_i+A_{0,i}\right)\right]^2+\fr12h_{ij}h^{ij}
+A_{0,i}A^{0,i}\right\}d^3x.
\label{3.6}
\end{eqnarray}
Then we get Dirac Hamiltonian which takes into account primary constraints 
\re{3.2} and \re{3.4}:
\beql{3.7}
H_{\rm D} = H_{\rm c}+\int\left[\lambda({\mit\Pi}+{\mit\Pi}^i n_i)
+\lambda_k((4\pi{\mit\Pi}^k+A_0{}^{,k})_{\perp}-n_jh^{jk})
\right]d^3x,
\eeq 
where $\lambda$, $\lambda_k$ are the Dirac multipliers.

The preservation of the constraint \re{3.2} in time produces 
the secondary constraint:
\begin{eqnarray}
&&
0 = \{{\mit\Pi}+{\mit\Pi}^i n_i, H_{\rm c} \} = -\rho - \partial_i
{\mit\Pi}^i_{\parallel}+\fr{\partial_i}{4\pi}\left[(A_0{}^{,i})_{\perp}
-n_jh^{ji}\right] \nn \\
&&
\approx -{\mit\Pi}^i{}_{,i} - \rho,
\label{3.8}
\end{eqnarray}
where $\approx$ means "weak equality'' in the sense of Dirac and
\beq
\rho(t, \bm x) = \sum\limits_{a=1}^{N} e_a \delta^3 (\bm x -
 \bm x_a(t))
\eeq
is a charge density.
 
Next, we consider commutation relations between constraints \re{3.2},
\re{3.4}, and \re{3.8}:
\begin{eqnarray}
&&
\{ {\mit\Pi}+{\mit\Pi}^i n_i, (4\pi{\mit\Pi}^k+A_0{}^{,k})_{\perp}
-n_jh^{jk} \}=0, \\
\label{3.9}
&&
\{ {\mit\Pi}^i{}_{,i} + \rho, (4\pi{\mit\Pi}^k+A_0{}^{,k})_{\perp}
-n_jh^{jk} \}=0, \\
\label{3.10}
&&
\{(4\pi{\mit\Pi}^k(t,\bm x)+A_0{}^{,k}(t,\bm x))_{\perp}-n_jh^{jk}(t,\bm x), 
\nn \\
&& 
(4\pi{\mit\Pi}^i(t,\bm y)+A_0{}^{,i}(t,\bm y))_{\perp}-n_jh^{ji}(t,\bm y) \}
\nn \\
&&
=-8\pi\left(\delta^{ki} - n^k n^i\right) 
\left(n^j \fr{\partial}{\partial x^j}\right) \delta^3 (\bm x-\bm y)\equiv
\Omega^{ki}(\bm x-\bm y).
\label{3.11}
\end{eqnarray}
We can check directly, that
\beql{3.12}
 \{ {\mit\Pi}^i{}_{,i} + \rho, H_{\rm c} \} = 0.
\eeq
Therefore, the two constraints
\beql{3.13}
\xi\equiv{\mit\Pi}+{\mit\Pi}^i n_i=0, \
\Gamma\equiv -{\mit\Pi}^i{}_{,i} - \rho\approx0,
\eeq
belong to the first class.
 
Taking into account \re{3.11}, we come to the conclusion that the constraints
\re{3.4} are of the second class. However, there are two independent 
second class constraints only. Then we can reduce them by means of Dirac
bracket:
\begin{eqnarray}
&&
\{F,G\}_{\rm D}=\{F,G\}-\int d^3xd^3y\{ F,(4\pi{\mit\Pi}^{\alpha}(t,\bm x)
+A_0{}^{,\alpha}(t,\bm x))_{\perp}-n_ih^{i\alpha}(t,\bm x)\} \nn \\
&&
\times
C_{\alpha\beta}(\bm x-\bm y) \{(4\pi{\mit\Pi}^{\beta}(t,\bm y)+
A_0{}^{,\beta}(t,\bm y))_{\perp}-n_jh^{j\beta}(t,\bm y), G \}, \
\alpha, \beta=1,2,
\label{3.14}
\end{eqnarray} 
where $C_{\alpha\beta}(\bm x-\bm y)$ is an inverse matrix to
 $\Omega^{\alpha\beta}(\bm x-\bm y)$:
\beql{3.15}
\int C_{\alpha\gamma}(\bm x-\bm z) \Omega^{\gamma\beta}(\bm z-\bm y) d^3z
=\delta_{\alpha}^{\beta}\delta^3 (\bm x-\bm y).
\eeq

In the next chapter we shall consider the elimination of the first class
constraints and the formulation of the Hamiltonian description in the terms of
independent physical variables.

Now let us canonically transform the field variables:
\begin{eqnarray}
&&
(A_0,\ {\mit\Pi},\ A_i,\ {\mit\Pi}_i) \mapsto (A_0,\ \ca E,\ \ca A_i,\
\ca E_i), \nn \\
&&
\ca E={\mit\Pi}+{\mit\Pi}^i n_i,\ \ca E_i={\mit\Pi}_i,\ \ca A_i=A_i+A_0 n_i.
\label{3.16}
\end{eqnarray} 
After the transformation the set of constraints of our system becomes
\begin{eqnarray}
&&
\ca E=0, \Gamma\equiv -\ca E^i{}_{,i} - \rho\approx0,
\label{3.17} \\
&&
4\pi \ca E^k_{\perp}-n_j(\ca A^{k,j}-\ca A^{j,k})=0.
\label{3.18}
\end{eqnarray}
Using \re{3.18}, we can rewrite the canonical Hamiltonian as
\begin{eqnarray}
&&
H_{\rm c} = - \sum\limits_{a=1}^{N}\left[
\fr{m_a^2+\bm k_a^2}{2\bm k_a \bm n}+e_aA_0(t,\bm x_a)\right] \nn \\
&&
+\int\left\{2\pi(n^i\ca E_i)^2-
\fr{1}{8\pi}\ca A_{i,j}(\ca A^{j,i}-\ca A^{i,j})
-A_0\ca E^i{}_{,i}\right\}d^3x,
\label{3.19}
\end{eqnarray}
 where now $k_{ai}=p_{ai}-e_a \ca A_i(t,\bm x_a)$.
  
\section{Elimination of the gauge degrees of freedom}
\setcounter{equation}{0}

 Let us consider the first class constraints. We see immediately that 
 $A_0, \ \ca E$ are a pair of conjugated gauge canonical variables. A second 
 such pair is formed by $Q = \Delta^{-1} \ca A^i{}_{,i}$ and $\Gamma$, 
 where $\Delta = \partial_i \partial_i, \ \Delta \Delta^{-1} = 1, \ 
 \Delta^{-1}_{\bm x} \delta^3(\bm x)=\Delta^{-1}(\bm x)\equiv
 -1/(4\pi|\bm x|)$.

 Therefore, we can separate the gauge degrees of freedom and gauge-invariant
 ones by means of the Hodge decomposition (see, e.g., \cite{gitman,lus}):
\beql{4.1}
 \ca A_i = \stackrel{\perp}{\ca A_i} + \partial_i Q, \ \
 \ca E_i = \stackrel{\perp}{\ca E_i} + \partial_i \Delta^{-1} (\Gamma + 
 \rho),
\eeq
 where
\beql{4.2}
\stackrel{\perp}{\ca A_i} = (\delta^j_i - \partial_i\Delta^{-1}\partial^j) 
\ca A_j, \
\stackrel{\perp}{\ca E_i} = (\delta^j_i - \partial_i\Delta^{-1}\partial^j) 
\ca E_j.
\eeq
 Then we have
\beql{4.3}
 \{ Q(t, \bm x), \Gamma(t, \bm y) \}= \delta^3 (\bm x-\bm y),
\ \{ \stackrel{\perp}{\ca A_i}(t, \bm x),
 \stackrel{\perp}{\ca E_j}(t, \bm y) \} =
 (\delta_{ij} - \partial^i \partial^j\Delta^{-1})\delta^3 (\bm x-\bm y).
\eeq

 Since $\stackrel{\perp}{\ca E^i{}_{,i}} = 0$, we can define 
 $\stackrel{\perp}{\ca E^i}$ as follows
\beql{4.4}
\stackrel{\perp}{\ca E^i} = \left(\delta_{\alpha}^i - 
\delta_3^i \fr{\partial_{\alpha}}{\partial_3} \right)
\fr{\tilde e^{\alpha}}{\sqrt{4\pi}}, \
\tilde e^{\alpha} = \sqrt{4\pi} \ca E^{\alpha}, \ \alpha = 1,2.
\eeq

 Now we have to do a canonical transformation to the new variables
 $((x^i_a, \tilde \pi_{ai})$, $(a_{\alpha}, \tilde e_{\alpha})$, $(Q, 
 \Gamma)$, $(A_0,\ca E))$, which is generated by the functional
\beq
F = \sum\limits_{a=1}^{N} x^i_a p_{ai} -
\int A_i \left[ \left( \delta^i_{\alpha} - 
\delta^i_3 \fr{\partial_{\alpha}}{\partial_3} \right)
\fr{\tilde e^{\alpha}}{\sqrt{4\pi}} + \partial^i \Delta^{-1}(\Gamma+\rho)
\right] d^3x.
\eeq
 We obtain
\begin{eqnarray}
&&
a_{\alpha}=-\fr{\delta F}{\delta e^{\alpha}}=\left(\delta^i_{\alpha} 
-\delta^i_3\fr{\partial_{\alpha}}{\partial_3} \right)\fr{\ca A_i} 
{\sqrt{4\pi}}, \\
\label{4.6}
&&
\tilde\pi_{ai}=\frp F{x^i_a}=p_{ai}-e_a\partial_i Q(\bm x_a).
\label{4.7}
\end{eqnarray}

 The transverse part of $\ca A_i$, is connected with the new canonical 
 variables as
\beql{4.8}
\stackrel{\perp}{\ca A_i} = \sqrt{4\pi}\left(\delta^{\alpha}_i -
\partial_i\Delta^{-1}\partial^{\alpha}\right) a_{\alpha},\ \ \alpha = 1,2.
\eeq

 Now it is convenient to take $\bm n=(0, 0, 1)$ and perform the 
 following canonical transformation of the field and particle momenta:
\beql{4.9}
\tilde e_{\alpha} = e_{\alpha}+\sqrt{4\pi}\partial_{\alpha}
\Delta^{-1} \rho, \
\tilde \pi_{ai} = \pi_{ai}-\sqrt{4\pi} e_a\partial_i\Delta^{-1}
\partial^{\alpha} a_{\alpha}(\bm x_a).
\eeq

 The set of constraints for our system in the terms of the new variables 
 can be written as follows:
\beql{4.10}
(\ca E, \ \Gamma, \ e_{\alpha}-a_{\alpha,3}) \approx 0.
\eeq

 We reduce the second class constraints by means of the Dirac 
 bracket:
\begin{eqnarray}
&&
\{F,G\}_{\rm D}=\{F,G\}-\int d^3xd^3y\{F,e_{\alpha}(t,\bm x)-a_{\alpha,3} 
(t,\bm x)\} \nn \\
&&
\times
C^{\alpha \beta}(\bm x-\bm y)\{e_{\beta}(t,\bm y)-a_{\beta,3} 
(t,\bm y),G\}.
\label{4.11}
\end{eqnarray}
 Here $\|C^{\alpha \beta}(\bm x-\bm y)\|$ is an inverse matrix to
\begin{eqnarray}
&&
{\bb W}=\|\{ e_{\alpha}(t,\bm x)-a_{\alpha,3}(t,\bm x),
e_{\beta}(t,\bm y)-a_{\beta,3}(t,\bm y)\}\| \nn \\
&&
=\left\|-2\delta_{\alpha \beta}\frp{}{x^3}\delta^3(\bm x-\bm y)\right\|, \\
\label{4.12}
&&
\int C^{\alpha \gamma}(\bm x-\bm z) W_{\gamma \beta}(\bm z-\bm y) d^3z 
=\delta^{\alpha}_{\beta}\delta^3(\bm x-\bm y).
\label{4.13}
\end{eqnarray}
 It is given by
\beql{4.14}
 C^{\alpha \beta}(\bm x-\bm y)=-\delta^{\alpha\beta} \delta(x^1-y^1) 
 \delta(x^2-y^2) {\rm sgn}(x^3-y^3),
\eeq
 because $\|C^{\alpha \beta}(\bm x-\bm y)\|$ must be antisymmetric.
 
 Let us introduce the following denotation:
\beql{4.15}
 \fr12\delta(x^1) \delta(x^2) {\rm sgn}(x^3)\equiv\fr{1}{\partial_3} 
 \delta^3(\bm x).
\eeq

 The final form of the Dirac bracket is
\begin{eqnarray}
&&
\{F,G\}_{\rm D}=\{F,G\}_{(x_a, \pi_a)}+\{F,G\}_{(A_0, \ca E)}+
\{F,G\}_{(Q, \Gamma)}+\fr12\{F,G\}_{(a, e)} \nn \\
&&
 +\fr12\int d^3x \left[
 \fr{\delta F}{\delta e_{\alpha}(\bm x)} \partial_3
 \fr{\delta G}{\delta e_{\alpha}(\bf x)} -
 \fr{\delta F}{\delta a_{\alpha}(\bf x)} \fr{1}{\partial_3}
 \fr{\delta F}{\delta a_{\alpha}(\bf x)}
 \right],
\label{4.16}
\end{eqnarray}
 where $\{F,G\}_{(x,\pi)}$ denote the standard Poisson bracket in the
 terms of $x$ and $\pi$.

 Elimination of constraints \re{4.10} into $H_{\rm c}$ leads to the physical
 Hamiltonian:
\begin{eqnarray}
&&
H_{\rm ph} = -\frac12 \sum\limits_{a=1}^{N} \left[
\pi_{a3} + \fr{(\pi_{a \alpha}-\sqrt{4\pi} e_a a_{\alpha}(\bm x_a))^2 +
m^2_a}{\pi_{a3}} \right]+\fr12\sum\limits_{a,b=1}^{N}\frac{e_a e_b}{|\bm 
x_a-\bm x_b|} \nn \\
&&
-\fr12\int\left(a_{\alpha}-\sqrt{4\pi} 
\fr{\partial_{\alpha}}{\partial_3}\Delta^{-1}\rho \right)
\Delta \left(a_{\alpha}-\sqrt{4\pi} 
\fr{\partial_{\alpha}}{\partial_3}\Delta^{-1}\rho \right)d^3x,
\label{4.17}
\end{eqnarray}
 which generates evolution of an arbitrary function $f$ depending on the
 gauge-inva\-ri\-ant variables $x^i_a$, $\pi_{ai}$ and $a_{\alpha}$, 
 $e_{\alpha}$ in the terms of the Dirac bracket \re{4.16}:
\beql{4.18}
\fr{{\rm d}f}{{\rm d}t}=\frp ft+\{f,H_{\rm ph}\}_{\rm D}.
\eeq

 The gauge-invariant volume element of the constrained field phase space
 $(a_{\alpha}, e^{\alpha})$ is written as
\beql{4.19}
 {\rm d}\Gamma^{\rm f}_{\rm ph}(t) =
\gamma \sqrt{{\rm Det}{\bb W}}
\prod\limits_{\alpha =1,2} \prod\limits_{\bm x}
\delta [e_{\alpha}(t, \bm x)-a_{\alpha,3}(t, \bm x)]
{\rm d}a_{\alpha}(t, \bm x) {\rm d}e_{\alpha}(t, \bm x),
\eeq
 where $\gamma$ is defined as a normalisation constant of the Gauss 
 integral \cite{slav}:
\begin{eqnarray}
&&
\gamma \int {\rm exp} \left( -\fr12
 \int a_{\alpha}(t, \bm x) L^{\alpha \beta}(t,\bm x,\bm y)
  a_{\beta}(t,\bm y)d^3xd^3y\right)
\prod\limits_{\alpha =1,2} \prod\limits_{\bm x}
 {\rm d}a_{\alpha}(t,\bm x) \nn \\
&& 
 = {\rm Det}^{-1/2} \| L^{\alpha \beta}(t,\bf x,\bm y) \|.
\label{4.20}
\end{eqnarray}

 Taking into account \re{4.19}, we can write the volume element of the
 physical phase space of the described system:
\beql{4.21}
 {\rm d}\Gamma_{\rm ph}(t) = {\rm d}\Gamma^{\rm p}_{\rm ph}(t)
 {\rm d}\Gamma^{\rm f}_{\rm ph}(t).
\eeq
 Here
\beql{4.22}
{\rm d}\Gamma^{\rm p}_{\rm ph}(t) = \prod\limits_{a=1}^{N} 
\prod\limits_{i=1}^{3} {\rm d}x^i_a(t){\rm d}\pi_{ia}(t).
\eeq
is the volume element of the particle phase space.
 
 Now we need check the Liouville theorem: $\Gamma_{\rm ph}(t)=
 \Gamma_{\rm ph}(t_0)$.
 
 It is well known from the classical mechanics, that system evolution in 
 the phase space can be described by means of a canonical transformation in 
 the terms of the Poisson bracket, which immediately leads to 
 the conservation of the phase space volume. This proves the 
 conservation of $\Gamma^{\rm p}_{\rm ph}$. 
 In our case the field evolution is generated by the 
 Dirac bracket. Nevertheless,  the conservation of 
 the volume of the constrained field phase space can
 be proved and we shall demonstrate its possible proof for the considered 
 Dirac bracket \re{4.16} in Appendix B.

 The volume element of the full phase space is
\beql{4.23}
{\rm d}\Gamma_{\rm full}(t)={\rm d}\Gamma_{\rm ph}(t){\rm d}\Gamma_{\rm 
g}(t),
\eeq  
 where ${\rm d}\Gamma_{\rm g}(t)$ is the gauge volume element of the phase 
 space:
\beql{4.24}
\prod\limits_{\bm x} \delta [\ca E(t, \bm x)]
\delta [\Gamma (t, \bm x)] {\rm d}A_0(t, \bm x) {\rm d}\ca E(t, \bm x)
{\rm d}Q(t, \bm x) {\rm d}\Gamma (t, \bm x).
\eeq
 However, we turn off ${\rm d}\Gamma_{\rm g}(t)$ from the further 
 description, because all thermodynamical characteristics of the system do 
 not depend on dynamics of the gauge degrees of freedom (see \cite{dn}).

\section{Statistical mechanics}
\setcounter{equation}{0}

 Now let us imagine that the physical initial data $x^i_a(t_0)$,
 $\pi_{ai}(t_0)$, $a_{\alpha}(t_0, \bm x)$, $e_{\alpha}(t_0, \bm x)$ are not 
 precisely known. Hence, we can introduce a probability density
 $\varrho (t_0) \equiv \varrho (t_0, \ x^i_a(t_0),\ \pi_{ai}(t_0), \ 
 a_{\alpha}(t_0, \bm x),\ e_{\alpha}(t_0, \bm x))$ for having 
 various initial states. This function satisfies the condition
\beql{5.1}
\int \varrho (t_0) {\rm d}\Gamma_{\rm ph}(t_0) = 1.
\eeq
Then average value of a general dynamical variable $f$ is defined by
\beq
\overline{f}(t) = \int f(t,\ x^i_a(t),\ \pi_{ai}(t),\ a_{\alpha}(t,\bm x),\
e_{\alpha}(t, \bm x)) \varrho (t) {\rm d}\Gamma_{\rm ph}(t).
\eeq

 Since $t_0$ has been randomly selected, we come to the relation
\beql{5.3}
1 = \int\varrho (t_0) {\rm d}\Gamma_{\rm ph}(t_0) = 
\int\varrho (t) {\rm d}\Gamma_{\rm ph}(t).
\eeq
 We have already seen that ${\rm d}\Gamma_{\rm ph}(t_0) =
 {\rm d}\Gamma_{\rm ph}(t)$, so one immediately obtains the Liouville 
 equation
\beql{5.4}
0 = \fr{{\rm d} \varrho (t)}{{\rm d} t} =
\fr{\partial \varrho (t)}{\partial t} +
\{ \varrho (t), H_{\rm ph}(t) \}_{\rm D}.
\eeq
 Taking into account \re{5.4}, we interpret $\varrho$ as an integral of 
 motion.

 Let us consider the canonical Gibbs ensemble in equilibrium. In this case 
 we have
\beql{5.5}
\varrho (t) = C {\rm e}^{-\beta H_{\rm ph}(t)},
\eeq
 where $\beta = 1/kT$ and $C$ is a normalisation constant.
 
 Partition function can be found as
\beql{5.6}
Z=\int\fr{e^{-\beta H_{\rm ph}(t)}}{(2\pi)^{3N} N!}{\rm d}\Gamma_{\rm ph}(t).
\eeq
 We shall find below the value of $Z$.

 Let us first rewrite the physical Hamiltonian as follows
\begin{eqnarray}
&&
H_{\rm ph} = -\fr12\sum\limits_{a=1}^{N}\left(\pi_{a3}+\fr{\pi^2_{a
\alpha}+m^2_a}{\pi_{a3}}\right)+\fr12\sum\limits_{a,b=1}^{N}\fr{e_a e_b} 
{|\bm x_a-\bm x_b|} \nn \\
&&
-2\pi\int\fr{\partial_{\alpha}}{\partial_3}\rho(t,\bm x)
\Delta^{-1}(\bm x-\bm y)
\fr{\partial_{\alpha}}{\partial_3}\rho(t,\bm y)d^3xd^3y \nn \\
&&
+2\pi\int\left(\fr{\partial_{\alpha}}{\partial_3}\rho(t,\bm x) 
+j^0_{\alpha}(t,\bm x)\right)G(t,\bm x-\bm y)
\left(\fr{\partial_{\alpha}}{\partial_3}\rho(t,\bm y)
+j^0_{\alpha}(t,\bm y)\right)d^3xd^3y \nn \\
&&
-\fr12\int\tilde a_{\alpha}(t,\bm x)\left(\Delta + 4\pi\sum\limits_{a=1}^{N}
\fr{e^2_a}{\pi_{a3}(t)} \delta^3 (\bm x-\bm x_a(t))\right)
 \tilde a_{\alpha}(t,\bm x) d^3x,
\label{5.7}
\end{eqnarray}
 where
\begin{eqnarray}
&&
\tilde a_{\alpha}(t,\bm x)=a_{\alpha}(t,\bm x)-\sqrt{4\pi}\int
G(t,\bm x-\bm y)\left(\fr{\partial_{\alpha}}{\partial_3}\rho(t,\bm y)
+j^0_{\alpha}(t,\bm y)\right) d^3y, \\
\label{5.8}
&&
j^0_{\alpha}(t,\bm x)=-\sum\limits_{a=1}^{N}e_a\fr{\pi_{a \alpha}(t)} 
{\pi_{a3}(t)}\delta^3(\bm x-\bm x_a(t)), \\
\label{5.9}
&&
\left[\Delta+4\pi\sum\limits_{a=1}^{N}\fr{e^2_a}{\pi_{a3}(t)}\delta^3
(\bm x-\bm x_a(t))\right] G(t,\bm x) = \delta^3(\bm x).
\label{5.10}
\end{eqnarray}

 Since $H_{\rm ph}$ does not depend on field momenta $e^{\alpha}$, after an 
 integration over $e^{\alpha}$ in \re{5.6} we obtain the expression for the 
 partition function:
\begin{eqnarray}
&&
Z = \gamma \sqrt{{\rm Det}{\bb W}}\fr1{N!}\int{\rm exp}(-\beta)\left[
-\fr12\sum\limits_{a=1}^{N}\left(\pi_{a3}+\fr{\pi^2_{a \alpha}+m^2_a} 
{\pi_{a3}}\right)+\fr12\sum\limits_{a,b=1}^{N}\fr{e_a e_b}{|\bm x_a 
-\bm x_b|}\right. \nn \\
&&
-2\pi\int\fr{\partial_{\alpha}}{\partial_3}\rho(t,\bm x)
\Delta^{-1}(\bm x-\bm y)
\fr{\partial_{\alpha}}{\partial_3}\rho(t,\bm y)d^3xd^3y \nn \\
&&
+2\pi\int\left(\fr{\partial_{\alpha}}{\partial_3}\rho(t,\bm x) 
+j^0_{\alpha}(t,\bm x)\right)G(t,\bm x-\bm y)
\left(\fr{\partial_{\alpha}}{\partial_3}\rho(t,\bm y)
+j^0_{\alpha}(t,\bm y)\right)d^3xd^3y \nn \\
&&
\left.
-\fr12\int\tilde a_{\alpha}(t,\bm x)\left(\Delta + 4\pi\sum\limits_{a=1}^{N}
\fr{e^2_a}{\pi_{a3}(t)} \delta^3 (\bm x-\bm x_a(t))\right)
\tilde a_{\alpha}(t,\bm x) d^3x\right] \nn \\
&&
\times
\prod\limits_{a=1}^{N} \prod\limits_{i=1}^{3}
\fr{{\rm d}x^i_a {\rm d}\pi_{ai}}{2\pi}
\prod\limits_{\alpha=1,2}\prod\limits_{\bm x}{\rm d}a_{\alpha}(t,\bm x)
\label{5.11}
\end{eqnarray}
 Here we can replace $da_{\alpha}$ by $d\tilde a_{\alpha}$. Taking into
 account \re{4.8}, the integration over $\tilde a_{\alpha}$ yields
\begin{eqnarray}
&&
Z = Z^{\rm f} \fr1{N!}\int\prod\limits_{a=1}^{N} \prod\limits_{i=1}^{3} 
\fr{{\rm d}x^i_a {\rm d}\pi_{ai}}{2\pi}{\rm exp}(-\beta)\left[
-\fr12\sum\limits_{a=1}^{N} \left(\pi_{a3}+\fr{\pi^2_{a 
\alpha}+m^2_a}{\pi_{a3}}\right)\right. \nn \\
&&
+\fr12\sum\limits_{a,b=1}^{N}\fr{e_a e_b}{|\bm x_a-\bm x_b|}
-2\pi\int\fr{\partial_{\alpha}}{\partial_3}\rho(t,\bm x)
\Delta^{-1}(\bm x-\bm y)
\fr{\partial_{\alpha}}{\partial_3}\rho(t,\bm y)d^3xd^3y \nn \\
&&
\left.
+2\pi\int\left(\fr{\partial_{\alpha}}{\partial_3}\rho(t,\bm x) 
+j^0_{\alpha}(t,\bm x)\right)G(t,\bm x-\bm y)
\left(\fr{\partial_{\alpha}}{\partial_3}\rho(t,\bm y)
+j^0_{\alpha}(t,\bm y)\right)d^3xd^3y\right] \nn \\
&&
\times {\rm Det}^{-1/2}\left\|\delta^{\alpha\beta}\left(\delta^3(\bm x-\bm 
y)+4\pi\Delta^{-1}(\bm x-\bm y)\sum\limits_{a=1}^{N}\fr{e^2_a}{\pi_{a3}}
\delta^3(\bm y-\bm x_a)\right)\right\|,
\label{5.12}
\end{eqnarray}
 where
\beql{5.13}
Z^{\rm f} = \fr{\sqrt{{\rm Det}{\bb W}}}
{\sqrt{{\rm Det}\|-\beta \delta^{\alpha\beta}\Delta\delta^3
(\bm x-\bm y)\|}}
\eeq
 represents the free field partition function.

\section{Conclusions}

In this paper we have once more demonstrated the usefulness of the
various forms of relativistic dynamics in treating the relativistic
particle systems. Constraint analysis of the considered problem shows
significantly different structures of Hamiltonian description of charged
particles, interacting by means of an electromagnetic field, in the
space-like and isotropic forms of dynamics. Specifically, the use of
front form of relativistic dynamics allows us to exclude the
electromagnetic field variables from classical partition function of the
system of charged particles. It is obvious that such an exclusion must
depend on the particular boundary conditions for the field variables,
but at present the proper sense of these conditions remains unclear.
The obtained representation for the partition function contains highly
nonlocal expressions, and their further analysis consist of a
complicated task. The consideration of the various approximation schemes
at this step seems to be inevitable.

It should be noted that exclusion of the field variables transform the 
problem into the domain of relativistic direct interaction theory (see, 
e.g., \cite{hoyle,gaida}). An application of such a theory to the 
consistent formulation of the relativistic statistical mechanics is just at 
the beginning.

On the other hand, the established Liouville equation may be used in
nonequilibrium situation as well. It serves an useful starting point for
deriving various new forms of kinetic equations for charged particle
system.

\section*{Acknowledgment}

We are greatly indebted to Yu.~Yaremko and V.~Shpytko for many
stimulating discussions. Numerous helpful conversation with A.~Duviryak
and his support in the performance of this work are especially
acknowledged.

\section*{Appendix A\protect\\
Relativistic free particle in an arbitrary form of dynamics}
\renewcommand{\theequation}{A.\arabic{equation}}
\setcounter{equation}{0}

The Lagrangian of relativistic free particle in a given form of dynamics
\re{2.5} is
\beq
L=-m\sqrt{(D\varphi)^2-\bm v^2}\equiv-m\Gamma^{-1}.
\eeq
The canonical momentum and Hamiltonian are given by
\beql{a2}
p_i=m\Gamma (g_{ij}v^j-\varphi_i\varphi_t),
\eeq
\beql{a3}
H=m\Gamma\varphi_tD\varphi,
\eeq
where the matrix $g_{ij}=\delta_{ij}-\varphi_i\varphi_j$ has been
introduced. The determinant of the matrix is
\beq
g\equiv\|g_{ij}\|=1-\bg {\varphi}^2,
\eeq
and $g\geq0$ as result of the condition \re{2.7b}.

Consider firstly the case $g>0$. Then the matrix $g_{ij}$ has an
inverse, $\tilde g_{ij}=\delta_{ij}+g^{-1}\varphi_i\varphi_j$ and from
\re{a2} it follows
\beq
\tilde g_{ij}p_ip_j=m^2\left(\Gamma ^2g^{-1}\varphi_t^2-1\right).
\eeq
Using \re{2.8}, we find
\beq
m\Gamma=\varphi_t^{-1}\sqrt{g(m^2+\bm p^2)+(p_i\varphi_i)}
\equiv\varphi_t^{-1}B,
\eeq
\beq
v_i=B^{-1}\varphi_t\left[p_i+g^{-1}\varphi_i(B+p_j\varphi^j)\right],
\eeq
\beq
D\varphi=B^{-1}\varphi_tg^{-1}\left(B+p_i\varphi^i\right).
\eeq
Combining these results with \re{a3}, one gets
\beql{a9}
H=\varphi_tg^{-1}
\left[\sqrt{g(m^2+\bm p^2)+(p_i\varphi^i)^2}+p_i\varphi^i\right].
\eeq
The case of isotropic forms of mechanics can be treated by taking the
limit $g\to+0$ and using that, in view of \re{a2},
\beq
p_i\varphi^i=m\Gamma\left[gv_i\varphi^i+\varphi_t(g-1)\right],
\eeq
so that $p_i\varphi^i\leq0$ as $g\to+0$. It gives
\beql{a11}
H=-\fr12\varphi_t\fr{m^2+\bm p^2}{p_i\varphi^i}.
\eeq

\section*{Appendix B\protect\\
Proof of the Liouville theorem for a given Dirac bracket}
\renewcommand{\theequation}{B.\arabic{equation}}
\setcounter{equation}{0}

 Let us show that evolution of the field variables as the generalized 
 canonical transformation in the terms of the Dirac bracket \cite{SudM} 
 conserves the volume of the constrained field phase space.

 We first consider transformation generated by some functional $G$ with
 an arbitrary parameter $\xi$:
\begin{eqnarray}
&&
\alpha_{\alpha} = a_{\alpha} + \xi \{ a_{\alpha}, G \}_{\rm D} =
a_{\alpha} + \fr{\xi}{2} \fr{\delta G}{\delta e^{\alpha}}
- \fr{\xi}{2} \fr{1}{\partial_3} \fr{\delta G}{\delta a^{\alpha}}, \nn \\
&&
\epsilon_{\alpha} = e_{\alpha} + \xi \{ e_{\alpha}, G \}_{\rm D} =
e_{\alpha} - \fr{\xi}{2} \fr{\delta G}{\delta a^{\alpha}}
+ \fr{\xi}{2} \partial_3 \fr{\delta G}{\delta e^{\alpha}}.
\label{b1}
\end{eqnarray}
 
 We immediately see that $e_{\alpha} - a_{\alpha,3} =
 \epsilon_{\alpha} - \alpha_{\alpha,3}$. So, the constancy
 condition of $\Gamma^{\rm f}_{\rm ph}$ leads to the relation:
\begin{eqnarray}
&&
\int \prod\limits_{\alpha =1,2} \prod\limits_{\bm x}
\delta [\epsilon_{\alpha}(t, \bm x)-\alpha_{\alpha,3}(t, \bm x)]
 {\rm d}\alpha_{\alpha}(t, \bm x) {\rm d}\epsilon_{\alpha}(t, \bm x) \nn \\
&&
 =\int J(\xi, 0) \prod\limits_{\alpha =1,2} \prod\limits_{\bm x}
\delta [e_{\alpha}(t, \bm x)-a_{\alpha,3}(t, \bm x)]
 {\rm d}a_{\alpha}(t, \bm x) {\rm d}e_{\alpha}(t, \bm x).
\label{b2}
\end{eqnarray}
 Here Jacobian $J(\xi,\eta)$ is defined by
\beql{b3}
J(\xi,\eta)\equiv{\rm Det}\left\| \fr{\delta (\alpha_{\alpha}(t,\bm x; \xi) 
\epsilon_{\alpha}(t, \bm x; \xi)}{\delta (\alpha_{\beta}(t, \bm y; \eta) 
\epsilon_{\beta}(t, \bm y; \eta)} \right\|.
\eeq
 If $\eta=0$, then we have
\beql{b4}
J(\xi, 0) = {\rm Det} \left\| 
\fr{\delta (\alpha_{\alpha}(t, \bm x; \xi) \epsilon_{\alpha}(t, \bm x; \xi)}
{\delta(a_{\beta}(t, \bm y) e_{\beta}(t, \bm y))} 
\right\|
\eeq
 It is evident that $J(0,0) = 1$. Let us compute
 $\left. ({\rm d}J(\xi, 0)/{\rm d}\xi)\right|_{\xi = 0}$. We get
\begin{eqnarray}
&&
\left. \fr{{\rm d}J(\xi, 0)}{{\rm d}\xi}\right|_{\xi = 0} =
\left. {\rm Tr}
\left\| \fr{\delta \partial_{\xi}\alpha_{\alpha}(t, \bf x; \xi)}
{\delta a_{\beta}(t, \bm y)} \right\| \right|_{\xi = 0} +
\left. {\rm Tr}
\left\| \fr{\delta \partial_{\xi}\epsilon_{\alpha}(t, \bm x; \xi)}
{\delta e_{\beta}(t, \bm y)} \right\| \right|_{\xi = 0} \nn \\
&&
 = {\rm Tr} \left\| \fr12 \fr{\delta^2 G}
{\delta a_{\beta}(t, \bm y) \delta e_{\alpha}(t, \bm x)} \right\| +
{\rm Tr} \left\| -\fr12 \fr{1}{\partial_3} \fr{\delta^2 G}
{\delta a_{\beta}(t, \bm y) \delta a_{\alpha}(t, \bm x)} \right\| \nn \\
&&
+ {\rm Tr} \left\| -\fr12 \fr{\delta^2 G}
{\delta e_{\beta}(t, \bm y) \delta a_{\alpha}(t, \bm x)} \right\| +
{\rm Tr} \left\| \fr12 \partial_3 \fr{\delta^2 G}
{\delta e_{\beta}(t, \bm y) \delta e_{\alpha}(t, \bm x)} \right\|.
\label{b5}
\end{eqnarray}
 It is obvious that the first and the third terms vanish. One of possible 
 ways demonstrating a cancellation of the second and fourth terms is 
 based on commutation of the Dirac bracket action and replacement of
 $a_{\alpha,3}$ by $e_{\alpha}$ (or $e_{\alpha}$ by $a_{\alpha,3}$ in view 
 of $e_{\alpha} - a_{\alpha,3} = 0$).
 
 Therefore, we have
\beql{b6}
\left. \fr{{\rm d}J(\xi, 0)}{{\rm d}\xi}\right|_{\xi = 0} = 0.
\eeq

 Since $J(\xi, 0) = J(\xi, \xi_1) J(\xi_1, 0)$, then
\beql{b7}
\left. \fr{{\rm d}J(\xi, 0)}{{\rm d}\xi}\right|_{\xi = \xi_1} =
\left. \fr{{\rm d}J(\xi, \xi_1)}{{\rm d}\xi}\right|_{\xi = \xi_1}
J(\xi_1, 0) = 0.
\eeq
Thus, $J(\xi, 0) = 1$ for all $\xi$.
 
 Now if we take $\xi=t$ and $G=H_{\rm ph}$, we come to conclusion 
\beq
\Gamma_{\rm ph}(t)=\Gamma_{\rm ph}(t_0),
\eeq
namely, the time evolution preserves the phase space volume 
$\Gamma_{\rm ph}$.

\end{document}